# Calculate the Optimum Threshold for Double Energy Detection Technique in Cognitive Radio Networks (CRNs)

**Morteza Alijani** [1]* , **Anas Osman** [2]

[1,2] Industrial Engineering Department, University of Trento, Italy

*Corresponding Author

## Abstract

One of the most important technical challenges when designing a Cognitive Radio Networks (CRNs) is spectrum sensing, which has responsibility for recognizing the presence or absence of the primary users in the frequency bands. A common technique used for spectrum sensing is double energy detection since it can operate without any prior information regarding the characteristics of the primary user signals. A double threshold energy detection algorithm is based on the use of two thresholds, to check the energy of the received signals and decided whether the spectrum is occupied or not. Furthermore, thresholds play a key role in the energy detection algorithm, by considering the stochastic features of noise in this model, as a result calculating the optimal threshold is a crucial task. In this paper, the Bi-Section algorithm was used to detect the optimum energy level in the fuzzy region which is an area between the low and high energy threshold. For this purpose, the decision threshold was determined by the use of the Bisection function for cognitive users. Numerical simulations show that the proposed method achieves better detection performance than the conventional double-threshold energy-sensing schemes. Moreover, the presented technique has advantages such as increasing the probability of detection ($p_d$) of primary users and decreasing the probability of Collison ($p_c$) between primary and secondary users.

## 1. Introduction

Recently, the market has seen a significant increase in demand towards the use of frequency spectrum, as a result, we face a shortage of radio frequency resources. Cognitive Radio (CR) has been introduced as the most prominent solution for managing this volume of user demand. This idea provides the possibility of using spectrum for secondary users when the primary users are offline [1]. Therefore, identifying the presence of primary users is the most important role in the implementation of CR. In fact, the efficiency and performance of this emerging technology depend on the function of spectrum sensing. The goal of this function is to determine the presence or absence of primary users. In other words, if the primary users are offline, the spectrum can be usable by the secondary users. One of the most influential parameters in spectrum sensing is accuracy. Therefore, many researchers have carried out studies in this filed and the outcome of their studies is the presentation of different techniques for spectrum sensing. In the following, Tab. 1 and Fig. 1, the summarization of the merits and drawbacks of these methods sensing are presented [2,3,4,19,20].

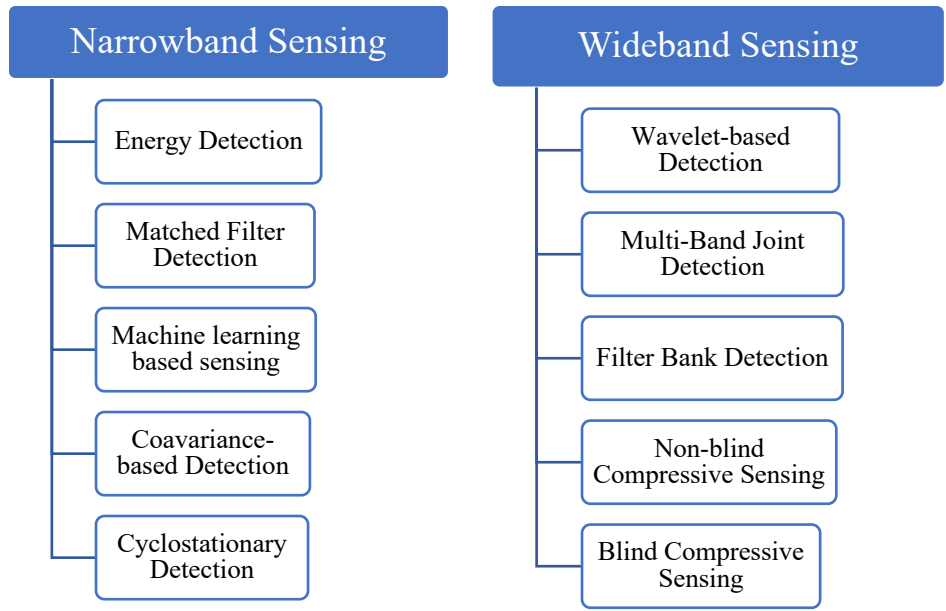

*Figure 1: Different methods for spectrum sensing*

According to Tab. 1, To check whether the primary user is present or not, in a matched filtering technique, information about the primary user signals is required, as well as, receivers for all types of signals. Furthermore, the power consumption and implementation complexity are too high. Whereas Cyclostationary detection algorithm provides reliable spectrum sensing, it has proven to be computationally complex and require long sensing time. However, Energy detection techniques are widely used for spectrum sensing, because it is simple to implement, does not require any prior information about the primary signal and has a low-cost implementation. In Ref [4], the authors provided a complete overview of the advantages and disadvantages of these methods, which can be referred to for further review [4]. Energy detection has been consistently used in comparison to other sensing methods. Many studies have been conducted to improve the performance of the energy detectors. Cooperative sensing with multiple CRs [1] and multi-band joint detection which evaluate each frequency band separately [2] are proposed to improve the energy detection performance. In order to detect the primary user, the optimum threshold should be selected. In this regard, double threshold energy detection and assigning the optimum threshold has been studied. Various double threshold methods [5-9] are proposed to improve the selection process. In addition to, an adaptive threshold setting algorithm is introduced for single-channel duty cycle estimation [9]. In [10], a sensing method using a double threshold for energy detection was proposed to reduce the communication traffic. In contrast, this method is applied in order to improve the macro detection capability of cognitive radio networks: assuming that the energy detector has two threshold values; each secondary user performs energy detection to sense the spectrum individually. Afterwards, the local decisions or observational values are reported to a Fusion Center (FC), and the FC will make a final decision to check if the primary user is absent or not. Simulation results in the latter research show that the spectrum sensing performance under AWGN channels has improved significantly.

The rest of the paper is organized as follows, in Section 2; the Energy Detection algorithm is introduced. In Section 3, the double threshold energy detection method is derived and the detection performance is analyzed. The proposed Bi-Section approach and simulation results are shown in Section 4, 5 respectively. Finally, the conclusions are drawn in Section 6.

*Table 1: Advantages and Disadvanatges of spectrum sensing techniques*

| Techniques | Advantages | Disadvanatges |
|---|---|---|
| **Matched Filter** | Optimum performance<br><br>Low computational cost<br><br>Short time to sense | Need to know information about the primary users<br><br>High power consumption<br><br>High complexity construction |
| **Energy Detection** | Low complexity<br><br>Independent of any prior information about the primary signal<br><br>Low computational cost | Vulnerable of unknown noise<br><br>The inability to distinguish between the original signal and the noise<br><br>Increasing the false alarm probability |
| **Cyclostationary Feature Detection** | Being robust against noise interference and uncertainty<br><br>The inability to distinguish between the original signal and the noise<br><br>Applicable for low signal to noise ratio | High complexity construction<br><br>Dependent of information about the primary users<br><br>Long time for sensing |
| **Covariance-based detection** | No prior knowledge of the primary user signal and<br><br>Noise is required<br><br>Blindly detection | Good computational complexity coming |
| **Machine learning based spectrum sensing** | Machine learning can detect if trained correctly can be a<br><br>Good approach<br><br>Minimize the delay of the detection<br><br>Use complex model in an easy manner | Complex techniques<br><br>Has to be adapted in learning in very fast changing environments<br><br>Features selection affects detection rate and adds complexity<br><br>High dataset has to be built |
| **Wavelet Multi-band joint detection** | Reduced latency compared to single band detection | Unaffordable sampling rate<br><br>High latency<br><br>High energy consumption<br><br>High complexity |
| **Filter bank** | Reduced latency compared to wavelet-based sensing techniques | High latency<br><br>High energy consumption<br><br>High complexity |

## 2. ENERGY DETECION

An energy detection algorithm means calculating the energy of the signal within a specific time and comparing it to a predetermined threshold value to make a decision regarding the presence of the primary user. Fig. 2, depicts the block diagram of the Energy detection algorithm. The first step in energy detection is to estimate the received power of the primary user $y(t)$.

In order to compute the power of the received signal, the output of the Bandpass filter of bandwidth $w$ is squared and integrated over an interval T. Finally, the integrated value is compared to a threshold $\lambda$ to decide the outcome.

Figure 2: Block Diagram of Energy Detection Algorithm

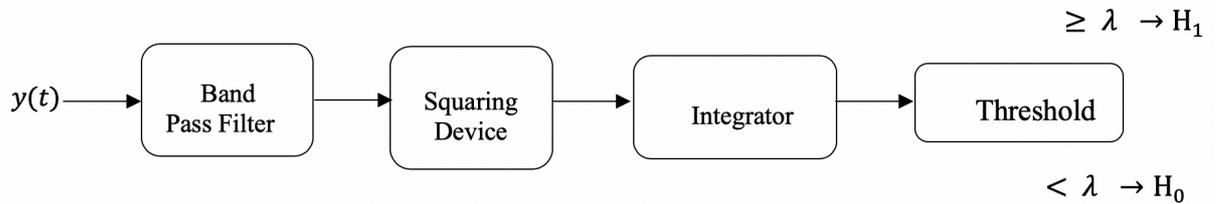

The threshold is a significant key parameter for the Energy detection method. It can be determined by the target performance metric, such as a false alarm or detection probabilities. Single threshold technique for energy detection shown in Fig. 3 and can be written as follows in (1) [11].

Figure 3: Single Treshold Technique

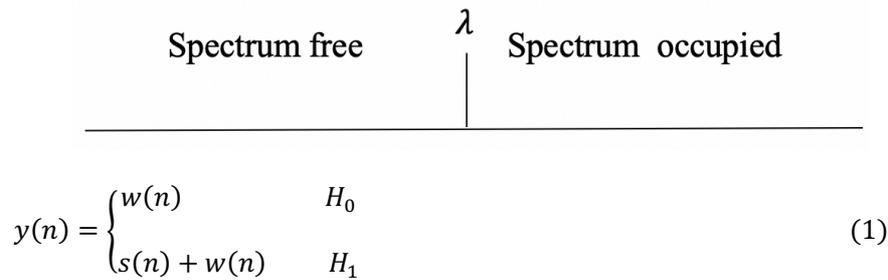

$$y(n) = \begin{cases} w(n) & H_0 \\ s(n) + w(n) & H_1 \end{cases} \quad (1)$$

where y(n) is the n-th sample of the received signal at the CR. $w(n)$ denotes the additive white Gaussian noise (AWGN) with zero mean and variance of $\sigma_w^2$. $s(n)$ represents the transmitted signal from the primary user with zero mean and variance of $\sigma_s^2$. $H_0$ is a label for the unoccupied spectrum and $H_1$ is a label for the occupied spectrum. Let $\gamma$ be the signal-to-noise ratio (SNR) which is defined as $\gamma = \sigma_s^2/\sigma_w^2$. Consider the CR employs energy detection with a single threshold that is shown in Fig. 2 and the test statistic (signal energy) is given by [7,12].

$$T_E(y) = \frac{1}{M}\sum_{n=1}^{M}|y(n)|^2 \quad (2)$$

where M is the total number of samples. For large M (M>10) [13], the test statistic $T_E(y)$ under two hypothesis approximately follows Gaussian distribution according to Central Limit Theorem (CTL) that is [14]:

$$T_E(y) \sim \begin{cases} N\left(\sigma_w^2, \frac{2}{M}\sigma_w^4\right) & H_0 \\ N\left((1+\gamma)\sigma_w^2, \frac{2}{M}(1+2\gamma)\sigma_w^4\right) & H_1 \end{cases} \quad (3)$$

For a given threshold $\lambda$, the primary user is estimated to be offline if $T_E(y) \leq \lambda$, otherwise the primary user is online. The probabilities of false alarm $P_f$ and detection $P_d$ can be computed as follows [14]:

$$P_f = P_r[T_E(y) > \lambda | H_0] = Q\left(\left(\frac{\lambda}{\sigma_w^2} - 1\right)\sqrt{\frac{M}{2}}\right) \quad (4)$$

$$P_d = P_r[T_E(y) > \lambda | H_1] = Q\left(\left(\frac{\lambda}{\sigma_w^2} - \gamma - 1\right)\sqrt{\frac{M}{2(2\gamma + 1)}}\right) \quad (5)$$

where $Q(.)$ is the complementary cumulative distribution function (CDF) of the variable standard Gaussian random i.e., $Q(x) = \left(\frac{1}{\sqrt{2\pi}}\right)\int_x^\infty \exp\left(\frac{-t^2}{2}\right)dt$, that can be written as follows: [11, 15]:

$$P_f = P_r[T_E(y) > \lambda | H_0] = \frac{\Gamma\left(u, \frac{\lambda}{2}\right)}{\Gamma(u)} \quad (6)$$

$$P_d = P_r[T_E(y) > \lambda | H_1] = Q_u(\sqrt{2\gamma}) \quad (7)$$

$$P_m = P_r[T_E(y) \leq \lambda | H_1] = 1 - Q_u(\sqrt{2\gamma}) = 1 - p_d \quad (8)$$

where $Q_u(a, b)$ is normalaized Marcum function with the order $u$. $\Gamma(a, b)$ is a non-complete gamma function; $\Gamma(a)$ is complete gamma function.

The probability of detection $(P_d)$ is the most concerning phase as it gives the correct sensing indication of the presence of primary users in the frequency band. The probability of miss-detection $(P_m)$ is just the complement of detection probability. Ultimately, the goal of the sensing schemes is to maximize the detection probability while minimizing the probability of false alarm $(P_f)$. However, there is always a trade-off between $P_d$ and $P_f$ probabilities. The Receiver Operating Characteristics Graph (ROC) presents valuable information about the detection probability versus false alarm probability ($P_d$ v/s $P_f$).

### 3. DOUBLE THRESHOLD ENERGY DETECION

In the double threshold energy detection algorithm shown in Fig. 4, there are two thresholds ($\lambda_L$ and $\lambda_H$). $E_K$ denotes the signal energy of the $k^{th}$ cognitive user. The $k^{th}$ user does not exist when $E_K$ is less than $\lambda_L$ and the decision is $H_0$. If $E_K$ is greater than $\lambda_H$, the $k^{th}$ cognitive user determines that the PU is online, correspondingly the decision is $H_1$. The decision is not clear nor robust, in the case that $E_K$ is between $\lambda_L$ and $\lambda_H$ (Fuzzy Region). Afterwards, the result is reported to FC in cooperative spectrum sensing processing.

*Figure 4. Double Threshold Energy Detection Algorithm*

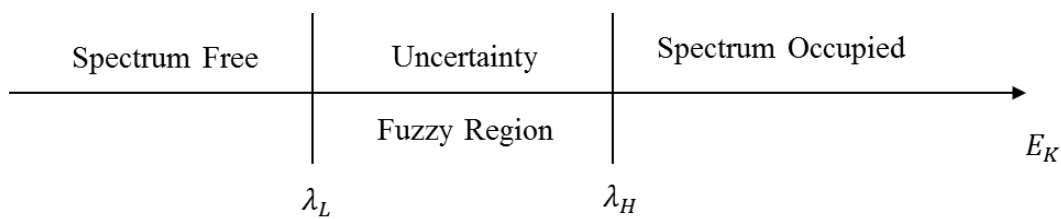

By considering Fig. 3, the single-threshold energy detection algorithm may cause critical interferences with the primary users. In order to alleviate interferences, the flowchart in Fig. 5 shows a Double threshold energy detection algorithm as an alternative approach [16].

*Figure 5. Flow Chart of Double Threshold Energy Detection Algorithm*

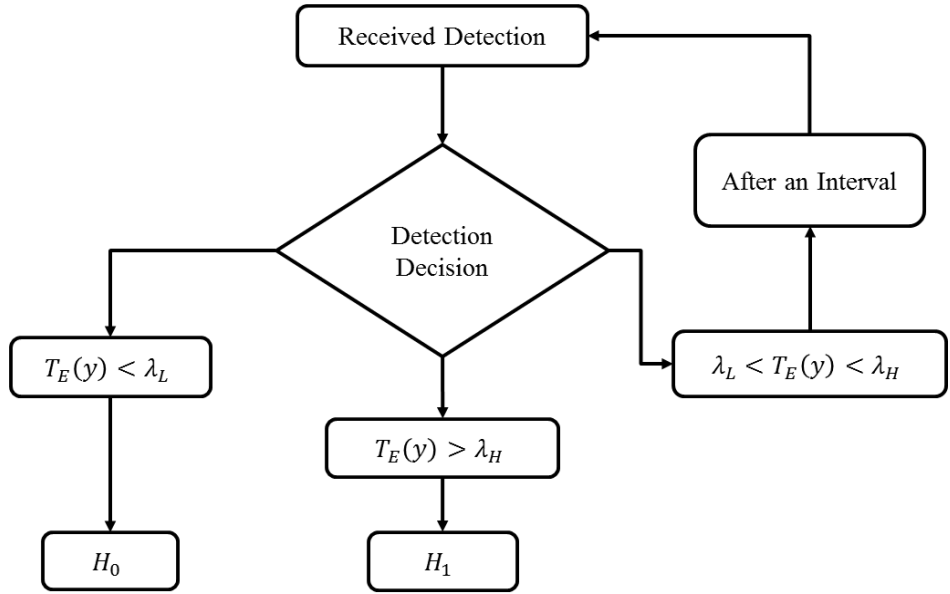

The performance of the double threshold energy detection method, the probability of detection, the probability of false alarm, and the probability of miss-detection can be computed as follows [16, 17, 18]:

$$P_f' = P_r[T_E(y) > \lambda_H | H_0] = \frac{\Gamma\left(u, \frac{\lambda_H}{2}\right)}{\Gamma(u)} \tag{9}$$

$$P_d' = P_r[T_E(y) > \lambda_H | H_1] = Q_u\left(\sqrt{2\gamma'}, \sqrt{\lambda_H}\right) \tag{10}$$

$$P_m' = P_r[T_E(y) \leq \lambda_L | H_1] = 1 - Q_u\left(\sqrt{2\gamma'}, \sqrt{\lambda_H}\right) \tag{11}$$

$$P_c = P_r([T_E'(y) < \lambda_L | H_1] \tag{12}$$

$$P_{na} = P_r([T_E'(y) > \lambda_L | H_0] \tag{13}$$

where $P_c$ is the probability of collision between the secondary user and primary user. $P_{na}$ is the probability of restricting the cognitive user from using the spectrum. In Double threshold energy detection method, if $E_K$ is between $\lambda_L$ and $\lambda_H$, the $k^{th}$ cognitive user cannot make a local decision, as a result, $E_K$ is sent to the DFC, resulting in an increment in the communication traffic, consumption of channel bandwidth, and consumption of energy in cognitive radio networks. To avoid the latter problems, an optimum double threshold detection with the aid of Bi-Section algorithm is proposed in section IV.

## 4. Bi-section Algorithm (Related Work)

In this section, the calculation of the optimum threshold using the Bi-Section algorithm is discussed. This method depends on $E_K$ because if $E_K$ falls in a fuzzy region, the decision becomes unreliable. When $E_K$ is between $\lambda_L$ and $\lambda_H$, the function $(f(x) = x - E_k)$ is defined to find the optimum threshold as follows:

---

**Bi-Section Algorithm: Calculating Optimal Threshold**

---

Set *tol* and *iter* to the tolerance and iterations respectively

Input lower ($\lambda_L$) and higher ($\lambda_H$) thresholds

**While** iter<maxiter && e>mintol **do**

  mid ← $(\lambda_L + \lambda_H)/2$

  % evaluate $f(x) = x - E_k$ at lower limit and midpoint

  $f(\lambda_L) \leftarrow f(low)$

$f(\lambda_H) \leftarrow f(mid)$

  **end**

  **if** $f(\lambda_L) * f(\lambda_H) > 0$ **then**

    % keep lower half of range: high=mid

    $f(\lambda_H) \leftarrow f(mid)$

  **end**

  **else**

    % keep lower half of range: low=mid

## 5. Numerical and simulation Results

In order to simulate the performance of the energy detector, ROC (Receiver Operating Characteristics, $(P_d/P_f)$ curve and complementary ROC $(P_m/P_f)$ was implemented through the use of Mont-Carlo technique. A random bit stream is multiplied by 1 MHz sinusoidal carrier signal to get 1 MHz BPSK modulated signal, which is transmitted in AWGN Channel with SNR $-14dB$ and u=5.

To evaluate the proposed approach in Tab. 2, the probability of detection in a double threshold energy method is compared to the probability of detection in optimum threshold. We define the following parameters $\lambda_L$=12, $\lambda_H$=18 and the probability of detection is $P_{d_2}$ =0.0676. Moreover, the probability of detection of an optimum threshold is assumed to be $P_{d_{opt}}$.

*Table 2: Comparison between the probability of detection in double and optimal threshold in energy detection technique.*

| $E_K$ | $\lambda_{opt}$ | $P_{dopt}$ | Improvement = $P_{dopt} - Pd_2$ |
|---|---|---|---|
| 12.5 | 12.37 | 0.2921 | 0.2245 |
| 14 | 13.87 | 0.2053 | 0.1377 |
| 15 | 17.62 | 0.0755 | 0.0079 |
| 15.5 | 15.37 | 0.1401 | 0.0725 |
| 16 | 16.12 | 0.1147 | 0.0471 |
| 16.5 | 17.62 | 0.0755 | 0.0079 |
| 17 | 16.87 | 0.0933 | 0.0257 |
| 17.5 | 17.62 | 0.0755 | 0.0079 |

In Tab. 2, the value of $(P_{dopt} - Pd_2)$ is regarded as an improvement to the proposed optimum threshold in comparison to double threshold energy detection that reached 0.2245 with $E_K$=12.5.

In Tab. 3, the probability of false alarm of the proposed optimum threshold is compared to the probability of false alarm of double threshold energy detection is discussed. The probability of a false alarm in the optimum threshold is $P_{fopt}$ and the probability of false alarm in a double threshold is $P_{f2}$=0.8735. Further investigation shows that the false alarm deterioration has increased to 0.0542.

*Table 3: Comparison between the probability of false alarm in double and optimal threshold in energy detection technique.*

| $E_K$ | $\lambda_{opt}$ | $P_{fopt}$ | $Deterioration = P_{fopt} - Pf_2$ |
|---|---|---|---|
| 12.5 | 12.37 | 0.9259 | 0.0542 |
| 14 | 13.87 | 0.9130 | 0.0395 |
| 15 | 17.62 | 0.8789 | 0.0054 |
| 15.5 | 15.37 | 0.8997 | 0.0262 |
| 16 | 16.12 | 0.8929 | 0.0194 |
| 16.5 | 17.62 | 0.8789 | 0.0054 |
| 17 | 16.87 | 0.8860 | 0.0125 |
| 17.5 | 17.62 | 0.8789 | 0.0054 |

Tab.4 represents the probability of miss-detection, which is valuable for performance analysis of the spectrum sensing scheme. The probability of miss-detection in the double threshold is assumed to be $P_{m2}$=0.9324 and the probability of miss-detection in the optimum threshold is $P_{m_{opt}}$.

*Table 4: Comparison between the probability of miss-detection in double and optimal threshold in energy detection technique.*

| $E_K$ | $\lambda_{opt}$ | $P_{mopt}$ | $Improvement = P_{mopt} - Pm_2$ |
|---|---|---|---|
| 12.5 | 12.37 | 0.7079 | − 0.2245 |
| 14 | 13.87 | 0.7947 | − 0.1377 |
| 15 | 17.62 | 0.9245 | − 0.0079 |
| 15.5 | 15.37 | 0.8599 | − 0.0725 |
| 16 | 16.12 | 0.8853 | − 0.0471 |
| 16.5 | 17.62 | 0.9245 | − 0.0079 |
| 17 | 16.87 | 0.9067 | − 0.0257 |
| 17.5 | 17.62 | 0.9245 | − 0.0079 |

One of the most important parameters during the analysis of spectrum sensing performance is the investigation of the probability of collision. Low collision probability indicates that the cognitive radio network is robust. To validate our approach we compute the probability of collision in varied scenarios of $\lambda_L$ and $\lambda_H$ for the double threshold method and the optimum threshold that can be shown in Tab. 5. $PC_2$ is the probability of collision in double threshold energy detection and $P_{Copt}$ is the probability of collision in an optimum threshold. In this table $E_k$=14.5.

*Table 5: Comparison between the probability of collision in double and optimal threshold in energy detection technique.*

| $\lambda_L$ | $\lambda_H$ | $\lambda_{opt}$ | $P_{C2}$ | $P_{Copt}$ | $P_f$ | Reduction $P_{Copt} - PC_2$ |
|---|---|---|---|---|---|---|
| 8 | 20 | 14.75 | 0.9627 | 0.8354 | 0.8561 | − 0.1273 |
| 7 | 22 | 21.06 | 0.9801 | 0.9731 | 0.8365 | − 0.0007 |
| 2 | 18 | 15.00 | 0.9324 | 0.8456 | 0.8753 | − 0.0868 |
| 11 | 26 | 13.81 | 0.9947 | 0.7917 | 0.7962 | − 0.2030 |
| 12 | 34 | 13.37 | 0.9997 | 0.7683 | 0.7146 | − 0.2314 |
| 5 | 21 | 14.00 | 0.9726 | 0.8012 | 0.8463 | − 0.1714 |
| 2 | 19 | 13.68 | 0.9496 | 0.7850 | 0.8658 | − 0.1646 |
| 10 | 21 | 14.81 | 0.9726 | 0.8379 | 0.8463 | − 0.1347 |

By considering Tab. 5 we can observe that the probability of collision has effectively decreased by using the proposed technique in comparison to the conventional technique. As a result, the interference between the primary users and secondary users has been decreased significantly. In Fig. 6 and Tab. 5, the comparison between single-double collision probabilities and optimum threshold are presented.

*Figure 6. Comparison between Single, Double Probability of Collision and Optimum Threshold*

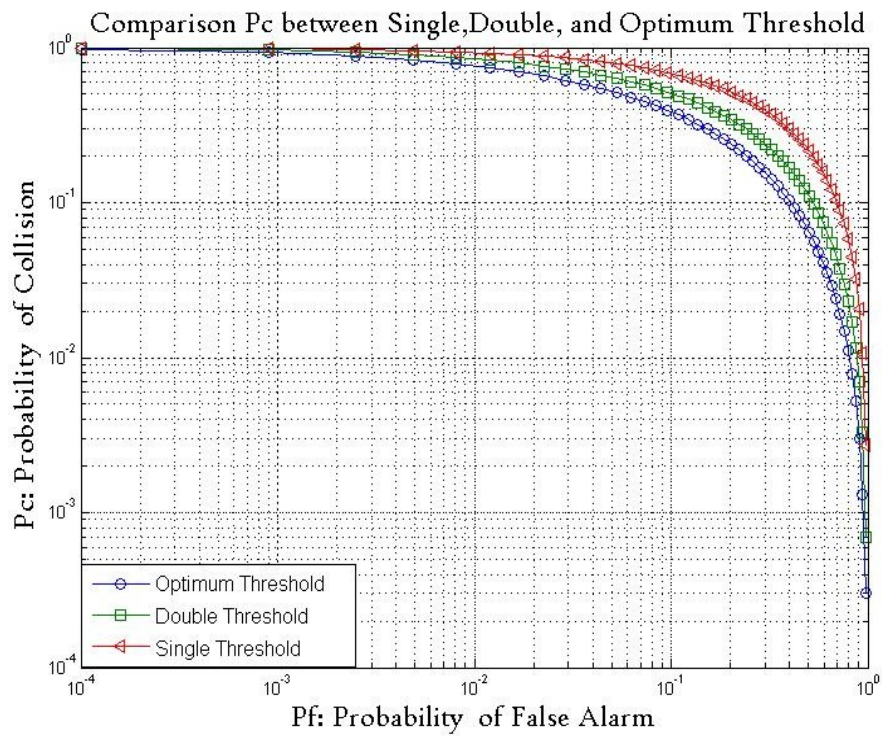

According to Fig.7, the probability of detection between the optimum threshold algorithm and the single-double threshold algorithm has improved.

*Figure 7. Comparison of ROC plot between Single, Double, and Optimum Threshold*

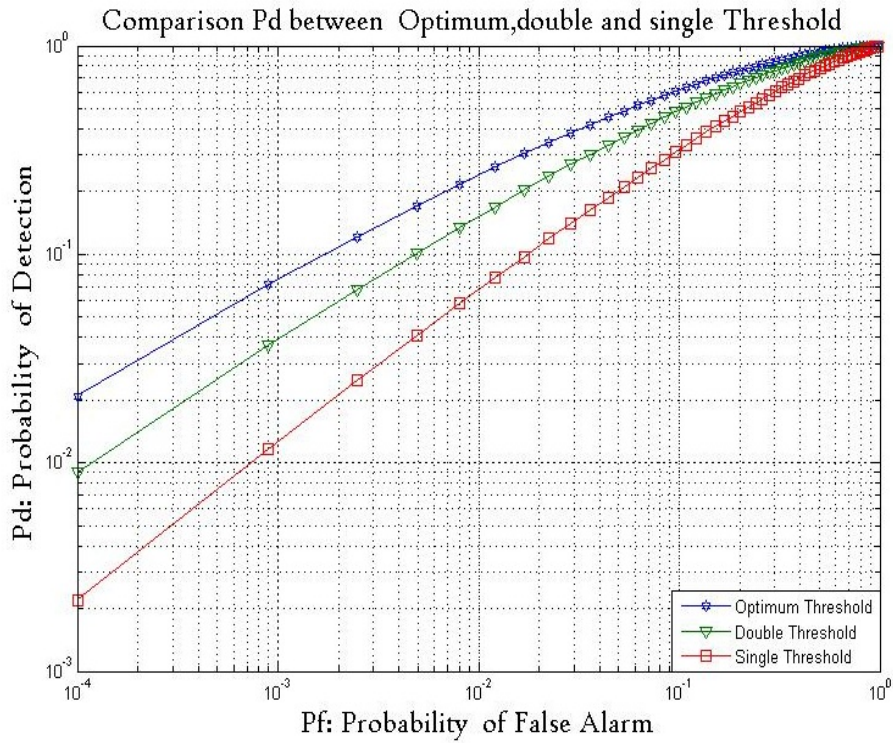

## 6. Conclusions

Spectrum sensing is one of the most crucial functions of Cognitive Radio Networks. In this paper, to achieve a robust double threshold energy detection algorithm that results in a decrease in communication traffic, channel bandwidth, and energy consumption. We have proposed a novel adaptive double threshold scheme for spectrum sensing based on energy detection. In this method, an optimal threshold in double energy detection is specified by applying the Bi-Section algorithm. To validate the proposed approach it's compared to the double threshold energy detection in Tables 2, 3, 4, 5, Fig. 6 and Fig.7 which depicts that the probability of detection and the probability of collision have been improved. On the contrary, by considering the trade-off between the probability of detection and false alarm, the probability of detection ($P_d$) has noticeably increased, however, the probability of false alarm ($p_f$) has increased slightly. Overall, the probability of collision ($p_c$) in the optimum threshold decreased in comparison with the conventional double energy detection algorithm, meaning that the proposed technique is robust.